\documentclass[runningheads]{llncs}

\usepackage[T1]{fontenc}
\usepackage{booktabs}
\usepackage{amsmath}
\usepackage{graphicx}
\usepackage{xcolor}
\usepackage{wrapfig}
\usepackage{subcaption}
\usepackage[most]{tcolorbox}
\usepackage[
    colorlinks=true,citecolor=blue,linkcolor=blue,urlcolor=blue,
    bookmarks=true, bookmarksopen=true, bookmarksdepth=3, bookmarksnumbered=true
]{hyperref}
\usepackage{color}

\usepackage{multirow}
\usepackage{array}
\usepackage{xcolor}
\usepackage{listings}
\usepackage{orcidlink}

\newtcolorbox{takeawaybox}{
  colback=blue!3,
  colframe=blue!60!black,
  boxrule=0.6pt,
  arc=2pt,
  left=4pt,
  right=4pt,
  top=4pt,
  bottom=4pt,
  before skip=6pt,
  after skip=6pt
}

\begin{document}
\title{A Measurement Study of Cryptographic Misuse in Embodied AI Mobile Applications}
\titlerunning{Cryptographic Misuse in Embodied AI Mobile Apps}

%
\author{ Junchao Li\inst{1}\orcidlink{0009-0005-3682-791X}
Xuelei Wang\inst{1}\orcidlink{0009-0006-6270-0952} \and
Yuhang Huang\inst{1}\orcidlink{0009-0008-2071-2108} \and
Qi Wang\inst{1}\orcidlink{0009-0006-7048-1884} \and
Boyang Ma\inst{1}\orcidlink{0000-0002-9849-4576} \and
Xuelong Dai\inst{1}\orcidlink{0000-0001-6646-6514}\and
Minghui Xu\inst{1}\orcidlink{0000-0003-3675-3461} \and
Yue Zhang\inst{1}\orcidlink{0000-0002-7786-0231}}


%
\authorrunning{X. Wang et al.}
%
\institute{
Shandong University, Qingdao, Shandong, China\\
}

\maketitle              
\begin{abstract}
Embodied AI (EAI) mobile applications are evolving from auxiliary user interfaces into active control-path components, directly linking mobile-side cryptographic security to cyber-physical trust. Despite this shift, existing security research predominantly focuses on embodied AI devices and cloud infrastructures, leaving the mobile control layer largely unexplored as a critical attack surface. To bridge this gap, we present the first large-scale measurement study of cryptographic misuse within the EAI mobile ecosystem. We construct EAIAppZoo, a benchmark of 507 real-world applications across six EAI domains, and employ an automated semantic-aware analysis pipeline to measure the prevalence and characteristics of five major cryptographic failure modes. Our measurement yields 12,975 misuse findings (with an evaluated precision of 80.74\%), revealing that these cryptographic failures are driven by EAI-specific engineering constraints rather than random developer errors. We uncover structural security trade-offs: latency-sensitive control paths systematically weaken transport protection, while the heavy reliance on offline device provisioning and legacy IoT SDKs exacerbates the local hardcoding of authentication credentials. Through real-world case studies, we demonstrate how these mobile-side cryptographic flaws bypass nominal network protections, enabling adversaries to intercept command channels and hijack the physical control of EAI entities. Ultimately, our findings highlight that mobile applications have become a fragile, yet overlooked, cryptographic trust boundary in cyber-physical systems.

\end{abstract}
%
\section{Introduction}

Embodied AI systems are rapidly moving from laboratory prototypes into real-world deployment scenarios, encompassing household cleaning robots, service robots, industrial robot arms, drones, wearable devices, and socially interactive platforms. In these ecosystems, mobile applications are no longer merely auxiliary user interfaces; they increasingly function as active control-path components responsible for device provisioning, remote command delivery, runtime monitoring, and cloud-assisted coordination. Through these roles, mobile applications directly mediate authentication and control instruction delivery. Unlike traditional mobile applications where a cryptographic compromise primarily leads to data privacy leakage, a cryptographic failure in an EAI control app translates directly into unauthorized physical actuation, kinetic damage, or unauthorized surveillance. This shifts the practical trust boundary of cyber-physical systems significantly towards the mobile layer. The mobile application layer serves as the critical bridge connecting users, cloud services, and physical entities, yet remains an insufficiently understood attack surface. In particular, the implementation and configuration of cryptographic logic along these complex EAI control paths remain poorly understood, with limited systematic analysis of how such mechanisms are deployed and potentially misused in practice.

Our preliminary observations indicate that cryptographic misuse in EAI mobile applications is not merely a collection of isolated developer errors, but rather driven by unique, systematic engineering constraints inherent to embodied systems. First, the necessity for \textit{local or offline control} (e.g., controlling a drone via direct Wi-Fi without internet access) creates significant friction with standard PKI infrastructures, often incentivizing developers to adopt insecure workarounds such as hardcoded static keys. Second, the strict \textit{low-latency requirements} for real-time video streaming and teleoperation frequently compel developers to intentionally downgrade or strip away transport-layer encryption (e.g., using plaintext RTSP or MQTT). Finally, the fragmented EAI supply chain heavily relies on \textit{legacy third-party IoT SDKs}, causing deprecated cryptographic primitives to spread virally across different device brands.

To systematically demystify this overlooked threat landscape, we present the first large-scale measurement study of cryptographic misuse specifically targeting the EAI mobile ecosystem. We aim to answer the following three core Research Questions (RQs):
\begin{itemize}
    \item \textbf{RQ1 (Prevalence \& Distribution):} 
    How prevalent are cryptographic misuses in the EAI mobile ecosystem, and how are they distributed across different application domains?
    \item \textbf{RQ2 (Framework Reliability):} 
    How reliable and accurate is the proposed semantic-aware analysis pipeline in identifying cryptographic misuses in real-world EAI mobile applications?
    \item \textbf{RQ3 (Cyber-Physical Impact):} 
    How do mobile-side cryptographic flaws translate into exploitable vulnerabilities that bypass network protections and hijack the physical control of EAI entities?
\end{itemize}

To answer these questions, we constructed \textbf{EAIAppZoo}, a benchmark dataset containing 507 real-world Android applications spanning six EAI categories. Instead of directly using off-the-shelf Android analysis tools, we developed an automated, semantic-aware static analysis pipeline. This pipeline seamlessly integrates dynamic memory unpacking with customized rule matching to precisely locate cryptographic misuses—including weak primitives, insecure parameter configurations, weak randomness, hardcoded secrets, and insecure communication practices—specifically along the practical control and authentication paths. 

Our measurement yields 12,975 misuse findings. We reveal that EAI mobile applications exhibit structured security trade-offs, exposing a fragile cryptographic trust boundary that severely undermines command authentication and transport integrity. Beyond large-scale prevalence, our case studies demonstrate that certain mobile-side cryptographic flaws can directly translate into practical cyber-physical attacks. In one representative case involving a Unitree Robotics quadruped robot application, a hardcoded cryptographic credential enabled the recovery of provisioning traffic, allowing attackers to reconstruct Wi-Fi credentials and pivot into the local network, thereby creating conditions for unauthorized command injection against the physical device.

In summary, we make the following key contributions:
\begin{itemize}
    \item \textbf{First Large-Scale Measurement on EAI Mobile Crypto-Security.} We construct EAIAppZoo and conduct the first systematic measurement of cryptographic misuses across 507 real-world EAI mobile applications, exposing the mobile layer as a critical vulnerability surface in cyber-physical ecosystems.
    \item \textbf{Demonstration of Cyber-Physical Attack Chains.} We bridge the gap between static code flaws and physical harm. Through in-depth case studies, we demonstrate how discovered mobile-side cryptographic vulnerabilities (e.g., hardcoded credentials and plaintext streaming) can be weaponized to intercept command channels and execute unauthorized physical hijacking of real-world EAI devices.
    \item \textbf{Revelation of EAI-Specific Root Causes.} We demystify the structural factors driving these widespread cryptographic failures, highlighting how low-latency operational constraints, offline provisioning requirements, and inherited legacy IoT SDKs systematically force dangerous security trade-offs.
\end{itemize}

The dataset is not publicly released at the current stage due to ongoing research considerations, and will be considered for public availability after publication.




\section{Responsible Disclosure}

Following responsible disclosure principles, we reported a subset of confirmed vulnerabilities to affected vendors whose products were involved in our empirical analysis. In particular, we submitted a detailed technical report to Unitree Robotics regarding a hardcoded cryptographic credential issue identified in one of its mobile application deployments, including the affected code patterns, potential unauthorized-access risks, and possible attack consequences.

Unitree Robotics actively acknowledged the reported issue and confirmed that the identified flaw could introduce unauthorized access risks under realistic deployment conditions. According to the vendor's response, the issue has entered its internal security review and remediation process, and we are closely working with the vendor to support mitigation and validation of the fix.

\section{Background}

\subsection{The Dual-Mode EAI Control Architecture}

\begin{wrapfigure}{r}{0.50\textwidth}
\vspace{-2mm}
\centering
\includegraphics[width=\linewidth]{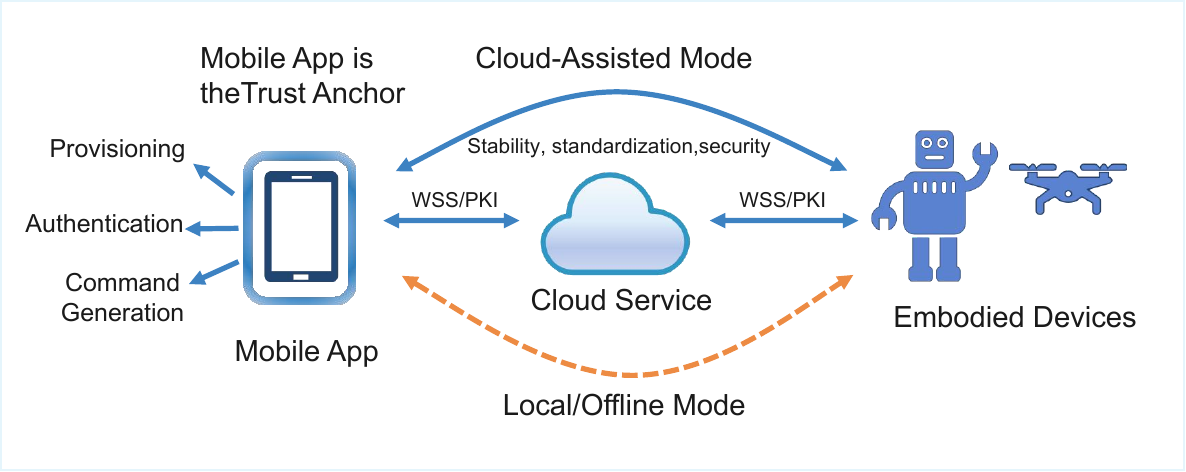}
\caption{Dual-mode control interactions in EAI systems.}
\label{fig:3arch}
\vspace{-2mm}
\end{wrapfigure}

A typical EAI ecosystem involves three tightly connected entities: the mobile application, the cloud service, and the physical embodied device, as illustrated in Fig.~\ref{fig:3arch}. Unlike conventional companion applications (which primarily act as passive data displays), mobile applications in EAI systems increasingly function as active, authoritative control-path bridges. They are directly responsible for device onboarding (provisioning), account authentication, real-time command generation, and firmware updates, effectively translating user intent into executable device actions. Crucially, EAI mobile control relies on a \textit{Dual-Mode Communication Architecture}, which enables operation across heterogeneous environments. The system supports both cloud-assisted communication for remote coordination and direct local communication for low-latency or offline control. These two modes coexist and are dynamically selected based on network conditions and operational requirements, forming a hybrid control paradigm distinct from traditional mobile-cloud interaction models.

\begin{itemize}
    \item \textbf{Cloud-Assisted Mode:} The application communicates with the device via cloud telemetry relay, typically utilizing standard HTTPS/WSS protocols secured by well-established Web Public Key Infrastructure.
    \item \textbf{Local/Offline Mode:} To ensure operational availability in environments without internet access (e.g., outdoor drone flights) or to minimize latency for real-time teleoperation, the application connects directly to the device via local networks (e.g., Wi-Fi AP mode, BLE, or local TCP/UDP). 
\end{itemize}

Within this dual-mode architecture, the mobile application often remains the primary trust anchor. It must locally manage credential storage, protocol initiation, and command construction before any physical actuation occurs.

\subsection{EAI vs. Traditional Robotics}

EAI systems differ fundamentally from traditional robotic systems in terms of architecture, control paradigm, and interaction model. Traditional robots are typically designed as closed, self-contained systems, where perception, planning, and control are tightly integrated within the device or a dedicated control unit. Their operation logic is relatively deterministic, task-specific, and often relies on pre-defined rules or classical control algorithms.  In contrast, EAI systems are inherently open and distributed. They integrate mobile applications, cloud services, and physical devices into a unified control loop, enabling more flexible and adaptive behaviors. Instead of relying solely on pre-programmed logic, EAI systems increasingly incorporate data-driven models, including large language models and multimodal perception modules, to interpret user intent and generate actions dynamically. Another key distinction lies in the control interface. Traditional robots are usually operated through specialized controllers or fixed interfaces, whereas EAI systems expose user-facing interfaces via mobile applications, supporting natural interaction modalities such as text, voice, and vision. This shift significantly lowers the barrier to interaction but also introduces new security risks, as the control surface becomes broader and more accessible. These architectural and operational differences are summarized in Table~\ref{tab:eai_vs_robot}.

\begin{table}[t]
\centering
\caption{Comparison between EAI Systems and Traditional Robotics}
\label{tab:eai_vs_robot}
\begin{tabular}{lrr}
\toprule
\textbf{Dimension} & \textbf{Traditional Robotics} & \textbf{EAI Systems} \\
\midrule
System Architecture & Closed, device-centric & Open, distributed (App--Cloud--Device) \\
Control Paradigm & Pre-defined, deterministic & Data-driven, adaptive \\
Interaction Interface & Dedicated controllers & Mobile apps (text/voice/vision) \\
Connectivity & Limited or optional & Always connected (cloud + local) \\
Intelligence Source & Embedded algorithms & AI models (e.g., LLMs, multimodal) \\
Control Path & Single or fixed & Dual-mode (cloud + local) \\
Update Mechanism & Infrequent, manual & Continuous, OTA updates \\
\bottomrule
\end{tabular}
\end{table}

\section{Motivation and Threat Model}

\subsection{Motivation}

Because mobile applications directly establish trust, cryptographic weaknesses at this layer critically undermine the entire cyber-physical system. To illustrate how such weaknesses emerge in practice, Listing~\ref{lst:motivating_example} presents a representative code pattern adapted from a real-world EAI mobile application (a quadruped robot companion app named \textit{JueYingLite2}), capturing three recurring classes of cryptographic misuse observed in our dataset.

\begin{lstlisting}[language=Java, caption={Real-world cryptographic misuses observed in an EAI mobile application named \textit{JueYingLite2}.}, label={lst:motivating_example}]
public void initAndPlay(String input) {
    try {
        // (1) Weak cryptographic primitive (MD5)
        MessageDigest md = MessageDigest.getInstance("md5");
        String hash = bytesToHex(md.digest(input.getBytes()));

        // (2) Insecure local communication (plaintext RTSP)
        String url = "rtsp://192.168.1.103:8554/camera";

        // (3) Hardcoded SDK credentials
        SmartPlayerJniV2 player = new SmartPlayerJniV2();
        player.SmartPlayerSetSDKClientKey(
            "8FC33AEE69AF4E7093FB95E9AFFE989C",
            "621E0DD5094841E2AB99C59C7173341A..."
        );
        player.SmartPlayerSetLowLatencyMode(1);
        player.SmartPlayerStartPlayback(url);
    } catch (Exception e) { e.printStackTrace(); }
}
\end{lstlisting}

Unlike conventional mobile applications where secure communication is typically enforced through standardized TLS-based APIs, EAI applications operate under a distinct set of engineering constraints that often correlate with degraded security practices. Based on our empirical observations, three recurring factors are closely associated with such cryptographic misuses:

\begin{itemize}
    \item \textbf{Use of Weak Cryptographic Primitives (Lines 3–5):}  
    The use of MD5 reflects reliance on outdated cryptographic primitives. Due to its lack of collision resistance, MD5 is unsuitable for security-sensitive operations, undermining \emph{data integrity guarantees}.
    
    \item \textbf{Latency-Sensitive Communication Paths (Lines 7-8):}  
    Real-time video streaming and teleoperation require low-latency communication. Applications may rely on plaintext RTSP streams over local IP addresses to reduce overhead, resulting in the absence of transport-layer protection and weakening \emph{communication confidentiality and integrity}.
    
    \item \textbf{Local Control Constraints and Credential Shortcuts (Lines 10–17):}  
    EAI applications frequently operate in local-network environments where standard PKI-based authentication is difficult to deploy. Developers therefore embed static credentials or SDK keys directly within the application to simplify device initialization and authorization, weakening \emph{credential security} and exposing secrets to reverse engineering.
\end{itemize}

Unlike traditional Android apps, EAI applications face unique design pressures—such as local connectivity, real-time demands, and proprietary SDKs—that drive recurring cryptographic misuses. This necessitates a large-scale empirical analysis to determine whether these vulnerabilities represent a systemic structural weakness across the ecosystem, or if they concentrate in specific domains tied to physical device control.

%
\subsection{Threat Model and Attack Surface}

In this measurement study, we assume a realistic, unprivileged attacker model targeting the mobile-side trust boundary.  We assume the cloud infrastructure and the internal hardware of the embodied device are trusted and secure. The attacker's capabilities are restricted to the following:

\begin{enumerate}
    \item \textbf{Reverse Engineer (APK Access):} The attacker can easily download the public EAI mobile application package from app markets and perform static reverse engineering to extract embedded assets, decompiled logic, and hardcoded cryptographic parameters.
    \item \textbf{Local Network Eavesdropper (MitM):} The attacker is situated on the same local network as the mobile device and the EAI entity (e.g., a shared Wi-Fi network, or within the broadcast range of the device's AP hotspot). The attacker can intercept, observe, and inject network traffic.
\end{enumerate}


\textbf{Attacker Goals \& Impact:} Unlike traditional attackers seeking data theft, the adversary in our EAI threat model aims for {Unauthorized Cyber-Physical Actuation}. By exploiting mobile-side cryptographic misuses (e.g., extracting a hardcoded session key to forge authentication tokens, or injecting malicious payloads into a plaintext control stream), the attacker seeks to bypass the nominal authorization mechanisms. The ultimate goal is to hijack the command channel, forcing the EAI device to execute unauthorized physical movements (e.g., altering a drone's flight path) or causing kinetic harm, completely independent of the legitimate user's control.

\section{Design of \textsf{EAGLE}}

\begin{figure}
\vspace{-2mm}
\centering
\includegraphics[width=\linewidth]{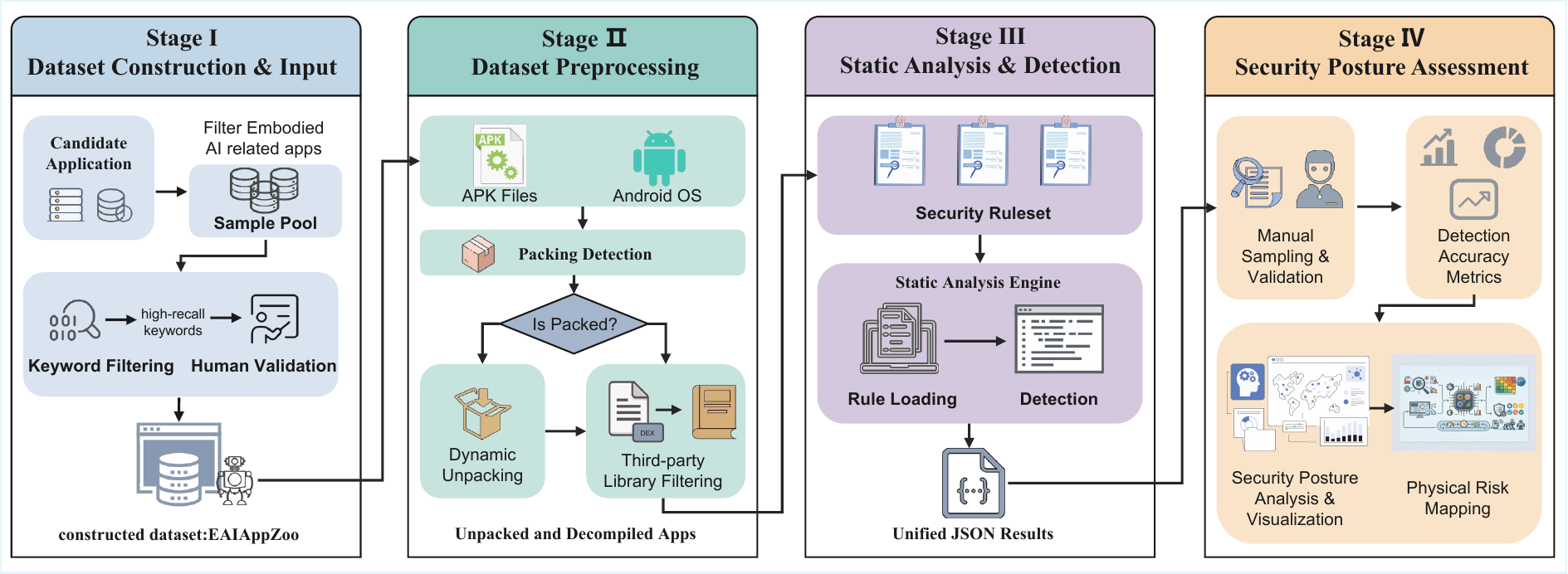}
\caption{Overview of the \textsf{EAGLE} framework.}
\label{fig:framework}
\vspace{-2mm}
\end{figure}

As illustrated in Fig.~\ref{fig:framework}, the proposed \textsf{EAGLE} (Embodied AI Global Leakage Evaluation) consists of four sequential stages: dataset construction, preprocessing, static detection, and security posture assessment.

\paragraph{\textbf{Stage I: Dataset Construction and Input.}}
The measurement pipeline begins with dataset construction, aiming to establish a representative sample of real-world embodied AI mobile applications before any security analysis is performed. Starting from candidate applications collected from mainstream app markets and open repositories, we first filter embodied AI related apps according to functional relevance, focusing on applications that directly participate in device control, telemetry, or local interaction with physical agents.

To improve recall, a keyword-based filtering strategy is applied using domain-relevant terms associated with drones, quadruped robots, educational robots, cleaning robots, and other embodied AI products. Because keyword filtering alone inevitably introduces noise, a manual validation step follows to remove irrelevant applications and confirm physical-control relevance. This process produces the final benchmark dataset, EAIAppZoo, which serves as the input for subsequent large-scale security measurement.

\paragraph{\textbf{Stage II: Dataset Preprocessing}}

To support automated large-scale analysis, each APK in EAIAppZoo is transformed into an analyzable intermediate representation through systematic preprocessing. We first unpack application packages and decompile DEX bytecode into Java-like source code using JADX~\cite{mauthe2021large}, while simultaneously preserving manifest metadata and package structures. Compared with raw bytecode, this representation provides higher semantic readability for tracking cryptographic API usage and credential logic.

Because EAI applications frequently rely on commercial protection mechanisms, the pipeline performs packing detection before analysis. If an application is identified as packed, dynamic unpacking is applied to recover executable code before further processing. After unpacking and decompilation, third-party libraries are aggressively filtered to reduce framework noise. Common generic dependencies (e.g., \texttt{androidx}, \texttt{okhttp3}, \texttt{retrofit2}) are excluded unless they directly interact with application-defined security logic. This ensures that the measurement reflects the security posture of EAI developers and proprietary SDK supply chains rather than generic Android framework behavior.

\paragraph{\textbf{Stage III: Static Analysis and Detection}}
Static analysis is driven by a rule-based semantic detection engine built on Semgrep. A custom security ruleset is loaded to identify cryptographic misuse patterns in decompiled source code. The rules target five major categories of security weaknesses frequently observed in mobile and IoT software: weak cryptographic primitives, insecure parameter configurations, weak randomness, hardcoded secrets, and plaintext or insecure communication practices.

For each application, the static analysis engine scans source files against the loaded ruleset and records all matched findings in a unified JSON format. Each finding preserves not only the matched code pattern itself, but also associated metadata such as file location, API context, and local surrounding code structures. This unified result format enables subsequent large-scale aggregation while preserving enough semantic detail for downstream interpretation.

The rules target five major categories of security weaknesses frequently observed in mobile and IoT software: weak cryptographic primitives, insecure parameter configurations, weak randomness, hardcoded secrets, and plaintext or insecure communication practices.

To improve methodological transparency, Table~\ref{tab:rule_inventory} further summarizes the operational rule inventory instantiated in the EAGLE pipeline. For each misuse category, we report the number of concrete detection rules together with their aligned CWE semantics, which serve as the standardized external reference for interpreting the security meaning of each matched finding.

\begin{table}[t]
\caption{Cryptographic misuse categories in EAGLE and example CWE mappings.}
\label{tab:rule_inventory}
\centering
\small
\renewcommand{\arraystretch}{1.4}
\begin{tabular}{p{3.5cm} p{5.6cm} p{2.3cm}}
\toprule
\textbf{Rule Category} & \textbf{Representative Description} & \textbf{Example CWE} \\
\midrule

\textbf{Weak Cryptographic Primitives} 
& \raggedright Deprecated or broken algorithms used in security-sensitive logic (e.g., MD5, DES, SHA-1). \cite{8863426,10.1145/2508859.2516693,sun2023cryptoeval} 
& CWE-327, CWE-328 \\

\textbf{Insecure Cryptographic Parameter Configurations} 
& \raggedright Unsafe parameter choices such as fixed IVs, ECB mode, or improper padding. \cite{ami2022crypto,chen2024towards,9529806} 
& CWE-323, CWE-326 \\

\textbf{Weak Randomness Generation} 
& \raggedright Predictable randomness sources or insecure pseudo-random generators in cryptographic operations. \cite{ami2022crypto,chen2024towards,sun2023cryptoeval}
& CWE-329, CWE-330 \\

\textbf{Hardcoded Keys or Sensitive Parameters} 
& \raggedright Static secrets embedded in code or assets, including keys, certificates, or authentication credentials. \cite{mykhaylova2024hardcoded}
& CWE-798, CWE-321 \\

\textbf{Insecure Communication Configurations} 
& \raggedright Plaintext transport or disabled certificate / hostname verification in communication channels. \cite{liu2018automatically}
& CWE-295, CWE-319 \\
\bottomrule
\end{tabular}
\end{table}

This rule inventory provides the operational basis for the subsequent large-scale scanning process, ensuring that ecosystem-level measurements remain both semantically grounded and reproducible across heterogeneous EAI application contexts.

\paragraph{\textbf{Stage IV: Security Posture Assessment}}
The final stage translates raw detection outputs into ecosystem-level security assessment. To estimate detection reliability, we perform manual sampling and validation on stratified subsets of findings, allowing us to compute detection accuracy metrics and calibrate the interpretation of static results.
At the ecosystem level, aggregated JSON outputs are used for security posture analysis and visualization. Both absolute misuse counts and normalized misuse densities are computed to avoid scale bias across applications of different sizes. This supports cross-category comparison of misuse prevalence and concentration.

To bridge static findings with embodied AI risk, findings located near communication interfaces, pairing logic, authentication modules, or control-relevant code paths are further interpreted under physical risk mapping. This contextualization allows the analysis to distinguish generic implementation flaws from security weaknesses that may directly weaken the mobile-device trust boundary and potentially affect physical actuation.

\section{The EAIAppZoo and Measurement Setup}

\subsection{EAIAppZoo Dataset Construction}

\begin{wrapfigure}{r}{0.50\textwidth}
\vspace{-2mm}
\centering
\includegraphics[width=\linewidth]{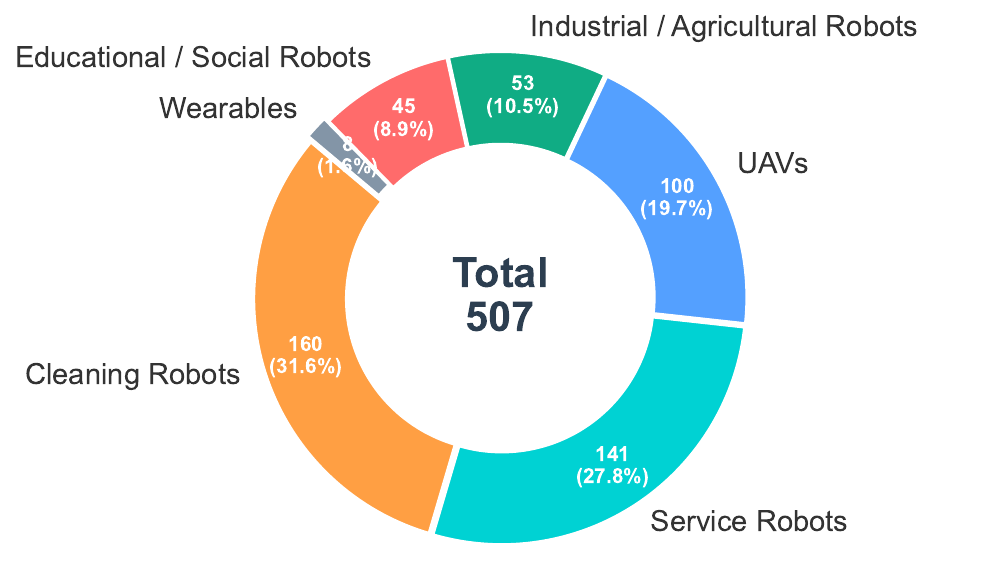}
\caption{Category distribution in EAIAppZoo.}
\label{fig:dataset}
\vspace{-2mm}
\end{wrapfigure}

To support the large-scale measurement of cryptographic misuses and answer our research questions, we require a highly representative, real-world dataset. To this end, we constructed \textbf{EAIAppZoo}, a benchmark dataset containing \textbf{507} in-the-wild Android applications. 

These applications were systematically collected from official application markets (e.g., Google Play) and the AndroZoo~\cite{alecci2024androzoo} repository. Candidate applications were initially retrieved using a comprehensive set of device-related keywords associated with embodied AI scenarios. However, to ensure the validity of our measurement regarding cyber-physical impact, we applied a strict manual filtering process. We excluded applications that serve merely as static manuals, companion guides, or cloud-only dashboards. Only applications explicitly designed to establish active control paths with physical devices—via local pairing, remote control, real-time monitoring, telemetry synchronization, or firmware updates—were retained.

As illustrated in Fig.~\ref{fig:dataset}, the finalized EAIAppZoo covers six distinct embodied AI domains: wearable devices, service robots, industrial and agricultural robots, educational and social robots, cleaning robots, and unmanned aerial vehicles (UAVs). This structural diversity enables us to observe domain-specific security trade-offs and variations in cryptographic implementations across the EAI ecosystem.

\subsection{Measurement Execution Environment}

To ensure consistency and eliminate manual bias across the large-scale measurement, all 507 applications within the EAIAppZoo were processed using our automated, semantic-aware analysis pipeline. Each application was evaluated independently through a standardized workflow: decompilation, third-party noise filtering, rule matching, and control-path contextualization.

The ecosystem-wide measurement was conducted on a high-performance workstation running Ubuntu 24.04.3 LTS, equipped with a 13th Gen Intel Core i9-13900HX CPU and 32 GB of memory, which provided sufficient computational resources for handling heavily obfuscated and large-scale APKs. The pipeline utilized JADX (v1.5.3) for reliable source-code reconstruction and Semgrep~\cite{bennett2024semgrep} (v1.146.0) as the core AST-matching engine. 

For each application, the pipeline generated a comprehensive security profile containing the total number of scanned files, distinct cryptographic misuses, and critically, the structural location of each finding (e.g., whether a hardcoded key was found within a local Wi-Fi pairing module). These structured outputs serve as the foundational data points for our subsequent statistical analysis and root-cause investigations.

\section{Measurement Results and Analysis}

To systematically address our research questions, this section presents the findings from our large-scale measurement of cryptographic misuses in EAI mobile applications. 

\subsection{Prevalence and Distribution of Misuses (RQ1)}

\begin{wrapfigure}{r}{0.45\textwidth}
\vspace{-2mm}
\centering
\includegraphics[width=\linewidth]{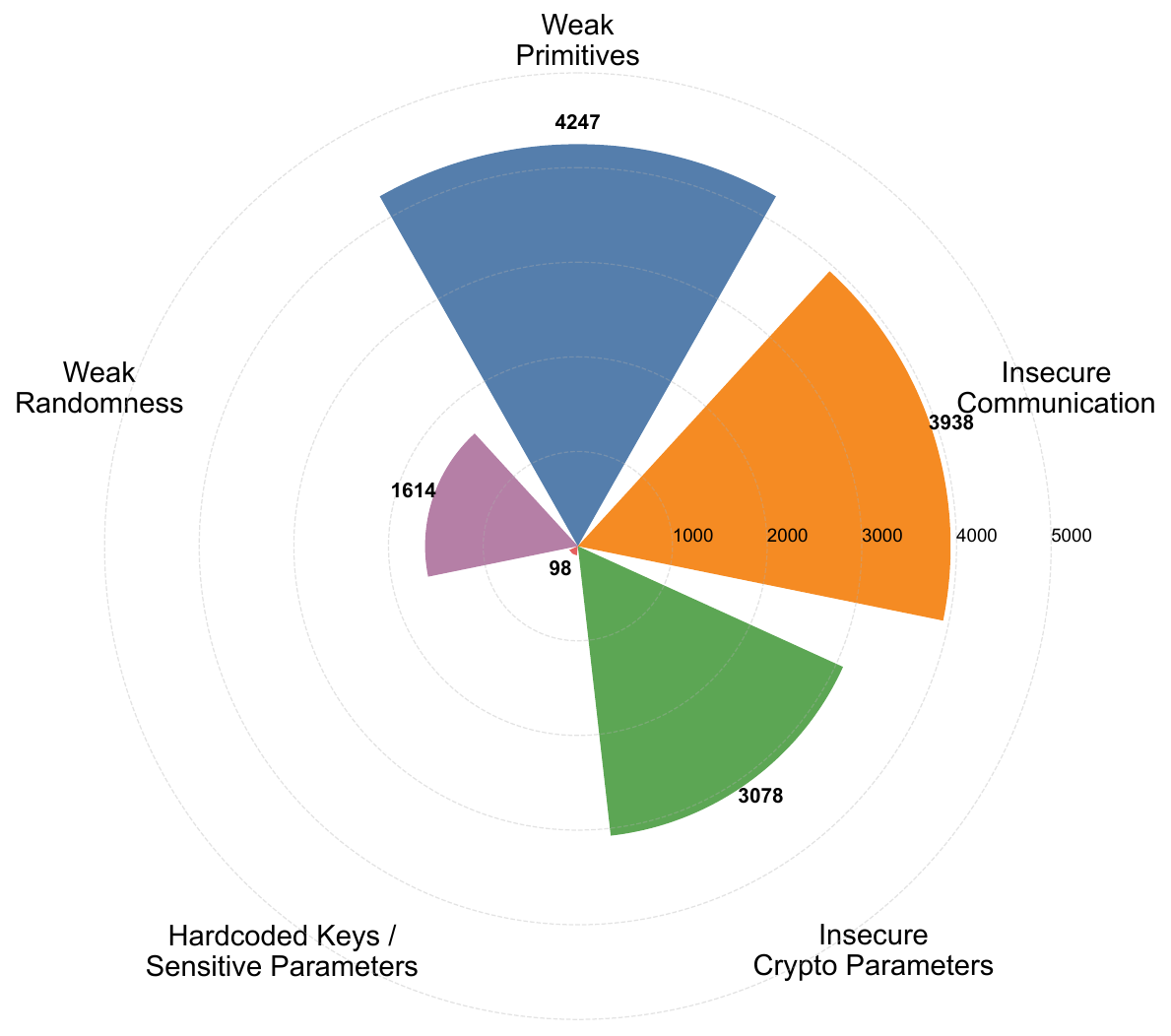}
\caption{Overall frequency of the five cryptographic misuse categories across the EAIAppZoo.}
\label{fig:misuse_category}
\vspace{-2mm}
\end{wrapfigure}

\textbf{Ecosystem-wide Prevalence.} 
Our measurement identifies 12,975 distinct cryptographic misuses across the 507 applications. As illustrated in Fig.~\ref{fig:misuse_category}, weak cryptographic primitives and insecure communication practices overwhelmingly dominate the EAI mobile landscape. The persistent use of deprecated algorithms (e.g., MD5, DES, and SHA-1 in security-sensitive control paths) and permissive transport configurations (e.g., disabled hostname verification, plaintext HTTP/TCP) indicates a severe and widespread degradation of basic trust mechanisms.

\textbf{Concentration Effects Across Applications.} 
Beyond aggregate prevalence, cryptographic misuses are not evenly dispersed across applications, but instead exhibit a clear concentration effect. A relatively small subset of applications contributes a disproportionately large number of misuse instances, suggesting the presence of high-risk clusters within the ecosystem. This skewed distribution implies that security weaknesses are not purely random implementation errors, but are instead influenced by shared development practices, common codebases, or domain-specific design patterns.

\begin{wrapfigure}{r}{0.60\textwidth}
\vspace{-2mm}
\centering
\includegraphics[width=\linewidth]{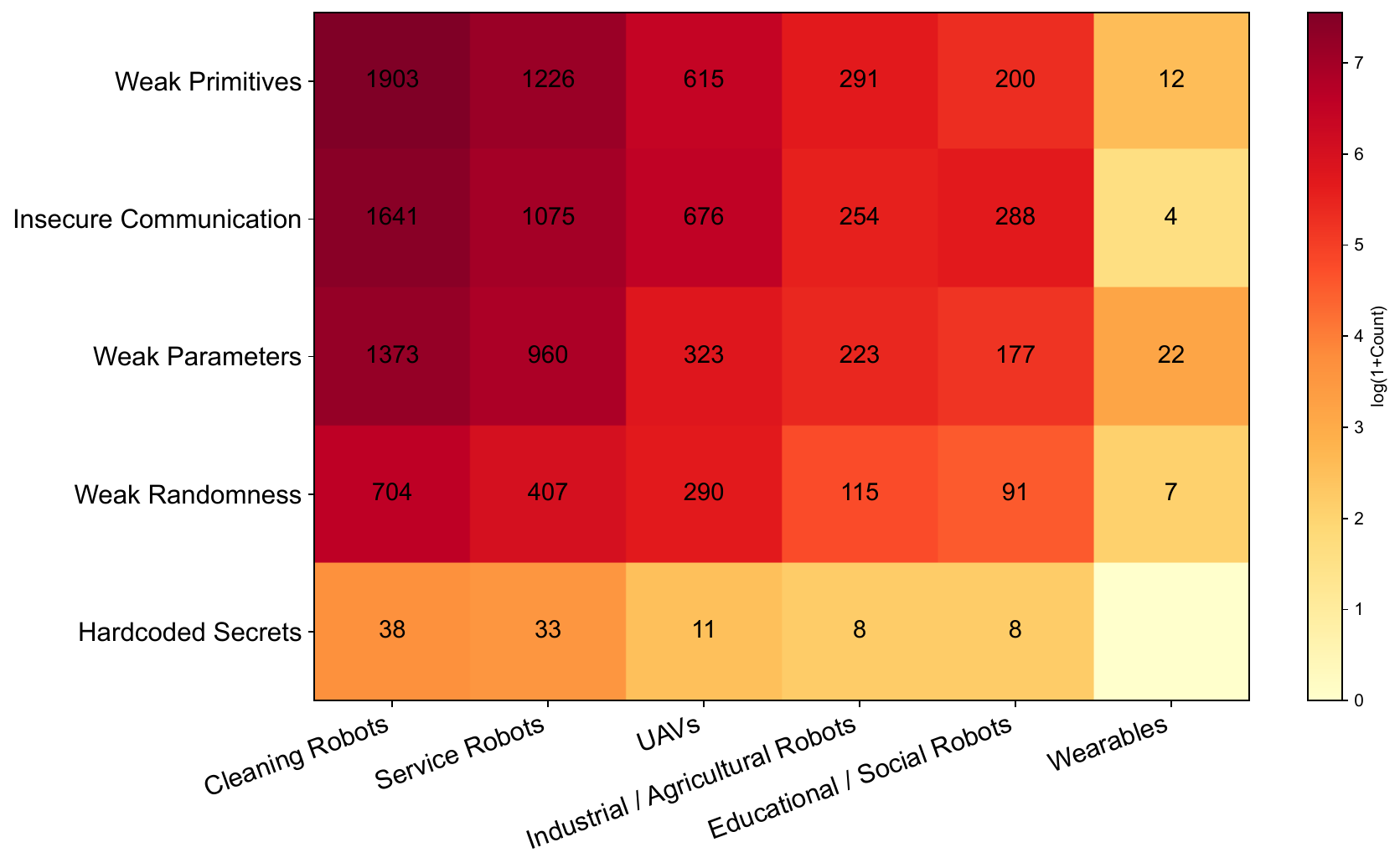}
\caption{Distribution of cryptographic misuse categories across EAI domains.}
\label{fig:heatmap}
\vspace{-2mm}
\end{wrapfigure}

\textbf{Domain-Specific Distribution Patterns.} 
To further understand how these risks are distributed across application categories, we analyze the normalized misuse composition across the six EAI domains. Fig.~\ref{fig:heatmap} presents the category-level distribution of misuse classes.

We observe that UAVs and Educational Robots are heavily associated with \emph{insecure communication} configurations, with a significant proportion of applications relying on plaintext channels for telemetry and command delivery. In contrast, Cleaning and Service Robots show a pronounced concentration of \emph{weak cryptographic primitives}, indicating that while encryption mechanisms are present, they often rely on outdated or insecure algorithms that provide limited effective protection.

\begin{takeawaybox}
\textbf{Takeaway for RQ1.} 
Cryptographic misuses are both widespread and structurally uneven across the EAI mobile ecosystem: they exhibit strong concentration effects at the application level and clear domain-dependent risk patterns, indicating that security weaknesses are shaped by both shared implementation practices and category-specific design choices.
\end{takeawaybox}

\subsection{Framework Reliability and Detection Precision (RQ2)}
\label{sec:validation}

To assess the reliability of the proposed semantic-aware analysis pipeline, we conducted a manual validation of detection results before deriving ecosystem-level conclusions. Because large-scale static analysis may introduce false positives under heterogeneous application contexts, especially when third-party libraries and obfuscated code coexist with application-specific logic, it is necessary to verify whether matched code patterns correspond to practically meaningful cryptographic misuses.

We adopted a stratified random sampling strategy across the six EAI application categories. In total, 244 control-relevant findings were manually inspected by security analysts to determine whether each matched instance reflected an actual cryptographic misuse under realistic code semantics.

The validation results are summarized in Table~\ref{tab:accuracy_stats}. Among the 244 sampled findings, 197 were confirmed as true positives, yielding an overall precision of 80.74\%, which indicates that the proposed pipeline achieves robust detection accuracy at ecosystem scale.

\begin{table}
\caption{Manual accuracy evaluation across six embodied AI categories.}
\label{tab:accuracy_stats}
\centering
\begin{tabular}{|l|l|l|l|}
\hline
\textbf{Application Category} & \textbf{Sampled Findings} & \textbf{TP} & \textbf{Precision} \\
\hline
Cleaning Robots & 87 & 75 & 86.21\% \\
Service Robots & 57 & 50 & 87.72\% \\
Unmanned Aerial Vehicles & 29 & 24 & 82.76\% \\
Industrial/Agricultural Robots & 14 & 10 & 71.43\% \\
Educational/Social Robots & 12 & 9 & 75.00\% \\
Wearable Devices & 45 & 29 & 64.44\% \\
\hline
\textbf{Total} & \textbf{244} & \textbf{197} & \textbf{80.74\%} \\
\hline
\end{tabular}
\end{table}

Detection precision remains highly stable across most domains. Wearable applications exhibit relatively lower precision (64.44\%), primarily because several findings originate from mature third-party DRM libraries whose internal cryptographic routines trigger static rules but are partially shielded by native-layer wrappers and aggressive obfuscation.

\begin{takeawaybox}
\textbf{Takeaway for RQ2.} 
The proposed semantic-aware analysis pipeline achieves robust practical precision (80.74\%) across heterogeneous EAI domains, indicating that it provides sufficient reliability to support ecosystem-level security measurement despite the presence of obfuscation and third-party library interference.
\end{takeawaybox}

\subsection{Cyber-Physical Impact and Exploitability (RQ3)}

To complement the macro-level statistics, we demonstrate how these mobile-side cryptographic misuses practically bypass nominal network protections and translate into physical threats. We highlight three representative real-world case studies extracted from the EAIAppZoo.

\textbf{Case 1: Rogue Device Takeover via Hardcoded PKI (mTLS Failure).} A service robot application was found to initialize a gRPC channel using embedded client-side certificate material. Although the communication stack nominally adopts mTLS, the client private key is packaged directly inside the APK, making the authentication boundary ineffective. The corresponding code excerpt is provided in Appendix~\ref{app:case1_code}.

\textit{Physical Impact.} An attacker can easily extract \texttt{client-key.pem} through lightweight reverse engineering. By loading the key into a custom script, the attacker can emulate a legitimate mobile client, bypass the mTLS trust boundary, and issue unauthorized physical actuation commands to the robot from within the network.

\textbf{Case 2: Kinetic Collision via Plaintext Command Injection.} A robotic arm control application was found to transmit MQTT commands over unencrypted TCP, exposing the physical command channel without transport-layer protection. The corresponding code excerpt is provided in Appendix~\ref{app:case2_code}.

\textit{Physical Impact.} Without transport protection, control payloads are broadcast in plaintext. A local Man-in-the-Middle attacker can not only eavesdrop on the robot's state, but also replay or inject forged commands (e.g., \texttt{\{"action":"rotate\_max"\}}), forcing unsafe movements that may lead to kinetic collisions and physical damage.


\textbf{Case 3: Full-Chain Compromise---From Hardcoded Keys to Home Network Breach and Kinetic Hijacking.}  
The most critical phase of EAI deployment is offline device provisioning, where the mobile app transmits the user's home Wi-Fi credentials to the robot via a local channel. To bypass dynamic key exchange complexities during this offline phase, developers often fallback to embedding static symmetric keys directly in the application bytecode. 

The static initialization logic of an \texttt{AESUtil} class extracted from a commercial quadruped robot application reveals that both the AES \texttt{secretKey} and the Initialization Vector (\texttt{IV}) are embedded as static constants in application code. The corresponding code excerpt is provided in Appendix~\ref{app:case3_code}.

This implementation is a textbook manifestation of \textbf{CWE-321 (Use of Hard-coded Cryptographic Key)}~\cite{cwe321}. Because these constants are identical across all application installations globally, an attacker can trivially extract them via static reverse engineering. Armed with the globally shared \texttt{secretKey} and static \texttt{IV}, the attacker can passively intercept and decrypt the entire provisioning session over the air. Table~\ref{tab:packet_analysis} illustrates the sequential decryption of the intercepted provisioning frames using the extracted key.

\begin{table}[htbp]
\caption{Decrypted provisioning frames exposing sensitive credentials.}
\label{tab:packet_analysis}
\centering
\begin{tabular}{|l|p{5.0cm}|p{4.5cm}|}
\hline
\textbf{Frame} & \textbf{Decrypted Payload (Hex)} & \textbf{Parsed Semantic Meaning} \\
\hline
706 & \texttt{52 0d 01 01 01 XX XX XX XX XX XX XX a2} & \texttt{cmd=SECRET, payload=[Redacted]} \\
746 & \texttt{52 04 07 a3} & \texttt{cmd=GET\_AP\_MAC} \\
816 & \texttt{52 04 02 a8} & \texttt{cmd=GET\_SN} \\
7778 & \texttt{52 05 03 02 a4} & \texttt{cmd=WIFI\_TYPE, type=0x02} (WPA2) \\
7804 & \texttt{52 0f 04 01 01 49 49 43 33 31 37 5f 35 47 4e} & \texttt{cmd=WIFI\_SSID, SSID="IIC317\_5G"} \\
7846 & \texttt{52 0e 05 01 01 69 69 63 31 32 33 34 35 65} & \texttt{cmd=WIFI\_PWD, PWD="iic12345"} \\
7893 & \texttt{52 07 06 43 4e 01 0f} & \texttt{cmd=COUNTRY, "CN", net\_open=1} \\
\hline
\end{tabular}
\end{table}

\textbf{Systemic and Cyber-Physical Impact:} This single cryptographic degradation effectively turns the EAI mobile app into a Trojan horse. An unauthenticated attacker outside the victim's house can sniff the encrypted packets, decrypt them, and immediately recover the plaintext WPA2 password of the victim's home network (Frame 7846). 

\begin{wrapfigure}{r}{0.5\textwidth}
\vspace{-2mm}
\centering
\includegraphics[width=\linewidth]{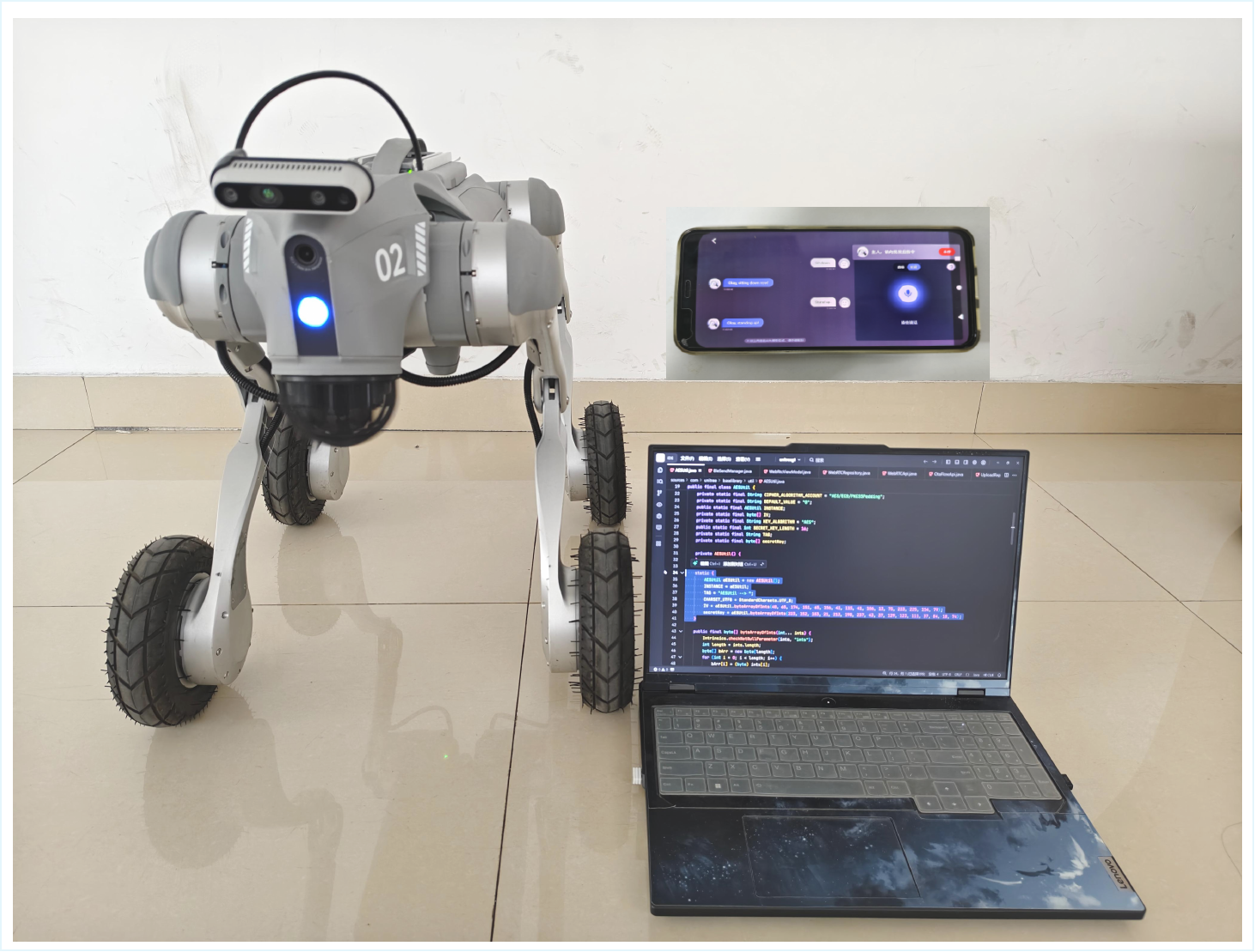}
\caption{Physical hijacking of a Unitree quadruped robot after command injection.}
\label{fig:robot_hijack}
\vspace{-2mm}
\end{wrapfigure}

Once the attacker joins the home Wi-Fi using the stolen credentials, they bypass all perimeter firewalls, allowing lateral movement to compromise other smart home devices. More dangerously, the attacker can reuse these hardcoded parameters to forge valid, encrypted kinetic control payloads. As illustrated in Fig.~\ref{fig:robot_hijack}, from within the local network, the attacker can inject forged motion commands, forcing the quadruped robot into unsafe physical states (e.g., triggering a sudden collapse or forcing it to stand). This successfully completes a full attack chain: transforming a static mobile credential flaw into a severe home network breach and immediate physical hijacking.

\begin{takeawaybox}
\textbf{Takeaway for RQ3.} 
Representative case studies show that mobile-side cryptographic misuses are not confined to software-layer weaknesses: once embedded into EAI control paths, they can directly undermine authentication boundaries, expose local trust anchors, and enable practical cyber-physical attacks ranging from unauthorized command injection to full physical hijacking of embodied devices.
\end{takeawaybox}

\section{Related Work}

\subsection{Security Research on Embodied AI Systems}

Recent security studies on embodied AI systems mainly focus on vulnerabilities in device-side architectures, model robustness, and system-level trustworthiness. Existing work has examined sensor robustness, actuator safety constraints, and sim-to-real deployment risks, showing that physical uncertainty and environmental disturbances may directly affect safe interaction in embodied environments~\cite{ji2025aialignmentcomprehensivesurvey,winfield_ethical_2025,8460875,kaushik2022safeapt}. At the model level, recent studies further show that embodied systems inherit jailbreak vulnerabilities from large language models, enabling malicious prompts to induce unsafe robotic behaviors or unintended physical actions~\cite{11128119,zhang_badrobot_2025,ravichandran2025safety}. Privacy leakage caused by multimodal sensing and unsafe data exposure has also become an important concern in embodied environments~\cite{nasr2023scalable,calo2011boundaries,chatzimichali2020privacysensitive}.

At the software and system level, prior work emphasizes trustworthy decision-making, legal accountability, and human--robot trust mechanisms~\cite{rachum2020whose,kok2020trust,esterwood2023three}. Recent system-oriented analyses argue that embodied AI security should be evaluated as an integrated cyber-physical system rather than isolated modules~\cite{tan2025towards,huang2025beyond}. However, existing studies still primarily concentrate on robot-side software stacks, middleware, and cloud infrastructures, while the mobile layer that directly mediates user control, authentication, and command delivery remains insufficiently examined.

\subsection{Cryptographic Misuse in Mobile Applications}

Cryptographic misuse has long been recognized as a persistent security problem in Android applications because incorrect cryptographic implementations directly weaken confidentiality, authentication, and secure communication. Prior studies report widespread issues such as hardcoded secrets, weak cryptographic primitives, insecure random number generation, and improper certificate validation across large Android ecosystems~\cite{10.1145/2508859.2516693,li2021static}. Subsequent large-scale analyses further show that cryptographic API misuse often originates from insecure code reuse, insufficient developer expertise, or incorrect API configuration~\cite{sun2023cryptoeval,mykhaylova2024hardcoded}. Recent work also explores automated repair and risk prioritization mechanisms to improve practical mitigation of cryptographic weaknesses~\cite{9529806,8863426}.

More broadly, mobile application security analysis commonly adopts static, dynamic, or hybrid approaches to detect vulnerabilities at scale~\cite{li2017static,qin2020vulnerability,bartel2014static,10.1145/2799979.2800036,10.1145/3321705.3329801,liu2018automatically}. However, most existing cryptographic misuse studies focus on conventional application domains such as finance, communication, or utilities, where security consequences remain largely digital. In embodied AI scenarios, mobile applications directly participate in device provisioning, command transmission, and cloud-device coordination, making mobile-side cryptographic weaknesses potentially more consequential because trust failures may propagate into cyber-physical interaction processes.

Overall, current literature has not systematically examined how cryptographic weaknesses manifest in embodied AI mobile applications, motivating the large-scale measurement in this work.

\section{Conclusion and Future Work}

This paper presents the first large-scale measurement study of cryptographic misuses within the EAI mobile ecosystem. By analyzing 507 real-world applications and identifying 12,975 security flaws, we reveal that mobile companion apps have become the most fragile cryptographic trust boundary in cyber-physical systems. Our findings demonstrate that widespread vulnerabilities—such as hardcoded symmetric keys and plaintext control channels—are not isolated coding errors, but systemic engineering compromises driven by low-latency demands, legacy SDK inheritance, and offline provisioning constraints. Ultimately, these mobile-side degradations allow adversaries to bypass nominal network defenses and execute unauthorized physical hijacking of EAI entities.

Future work will expand upon this static ecosystem measurement by incorporating dynamic execution analysis and physical testbed validations. This will allow us to further systematically map how these mobile-layer cryptographic failures propagate into complex cyber-physical attack chains in real-world deployment environments.

\begin{credits}
\subsubsection{\ackname}
The authors would like to thank the anonymous reviewers for their valuable comments and suggestions.

\subsubsection{\discintname}
The authors have no competing interests to declare that are relevant to the content of this article.
\end{credits}

\bibliographystyle{splncs04}
\bibliography{ref}

\appendix
\section{Representative Code Snippets}
\label{app:code_examples}

\subsection{Case 1: Embedded Client Private Key in mTLS Initialization}
\label{app:case1_code}

Listing~\ref{lst:mTLS} shows the embedded client private key used during gRPC channel initialization.

\begin{lstlisting}[language=Java, caption={Embedded client private key rendering mTLS authentication ineffective.}, label={lst:mTLS}]
private final ManagedChannel newChannel(String host, int port) throws IOException {
    Application app = AppUtils.INSTANCE.getApp();
    InputStream ca = app.getAssets().open("cert/ca-cert.pem");
    InputStream cert = app.getAssets().open("cert/client-cert.pem");
    InputStream key = app.getAssets().open("cert/client-key.pem");
    // ...
}
\end{lstlisting}

\subsection{Case 2: Plaintext MQTT Command Transmission}
\label{app:case2_code}

Listing~\ref{lst:MQTT} shows the plaintext MQTT initialization logic extracted from the robotic arm control application.

\begin{lstlisting}[language=Java, caption={Plaintext MQTT communication exposing the physical command channel.}, label={lst:MQTT}]
public class MQTTService extends Service {
    public static final String baseHost = "192.168.1.103:1883";
    public void init() {
        this.host = "tcp://" + baseHost;
        MqttAndroidClient client = new MqttAndroidClient(this, this.host, this.clientId);
    }
}
\end{lstlisting}

\subsection{Case 3: Hardcoded AES Parameters in Offline Provisioning}
\label{app:case3_code}

Listing~\ref{lst:AES} shows the static AES initialization logic extracted from the quadruped robot control application. Sensitive byte positions are partially masked for responsible disclosure.

\begin{lstlisting}[language=Java, caption={Hardcoded AES key and static IV in a quadruped robot control application.}, label={lst:AES}]
static {
    AESUtil aESUtil = new AESUtil();
    INSTANCE = aESUtil;
    TAG = "AESUtil --> ";
    CHARSET_UTF8 = StandardCharsets.UTF_8;
    IV = aESUtil.byteArrayOfInts(40, 65, XX, XX, XX, XX, XX, XX, XX, XX, XX, XX, XX, XX, 154, 79);
    secretKey = aESUtil.byteArrayOfInts(223, 152, XX, XX, XX, XX, XX, XX, XX, XX, XX, XX, XX, XX, 18, 74);
}
\end{lstlisting}

\end{document}